\documentclass[sigconf]{acmart}
\AtBeginDocument{%
  }

\usepackage{enumitem}
\usepackage{bm}
\usepackage{booktabs}
\usepackage{xcolor}
\usepackage{colortbl}

\usepackage{fancyvrb}

\setcopyright{acmlicensed}
\copyrightyear{2026}
\acmYear{2026}
\acmDOI{XXXXXXX.XXXXXXX}
\acmConference[UMAP '26]{34th ACM International Conference on User Modeling, Adaptation and Personalization}{June 08--11, 2026}{Gothenburg, Sweden}

\begin{document}

\pagenumbering{gobble}
\setcounter{page}{0}

\section*{ACM Copyright Statement}

\noindent
Copyright \textcopyright 2026 held by the owner/author(s). Publication rights licensed to ACM. Permission to make digital or hard copies of all or part of this work for personal or classroom use is granted without fee provided that copies are not made or distributed for profit or commercial advantage and that copies bear this notice and the full citation on the first page. Copyrights for components of this work owned by others than the author(s) must be honored. Abstracting with credit is permitted. To copy otherwise, or republish, to post on servers or to redistribute to lists, requires prior specific permission and/or a fee. Request permissions from permissions@acm.org.

\vspace{3em}
\noindent\textbf{Accepted to be published in:} Proceedings of the 34th ACM International Conference on User Modeling, Adaptation and Personalization, 2026 (\textit{UMAP'26, Gothenburg, Sweden})

\vspace{3em}
\noindent\textbf{Cite as:}
\vspace{0.5em}

\noindent
Rafael T. Sereicikas, Pedro R. Pires, Gregorio F. Azevedo, and Tiago A. Almeida. 2026. Learning Behaviorally Grounded Item Embeddings via Personalized Temporal Contexts. In \textit{Proceedings of the 34th ACM International Conference on User Modeling, Adaptation and Personalization} (Gothenburg, Sweden) \textit{(UMAP'26)}. Association for Computing Machinery, New York, NY, USA, 1-9, doi:10.1145/3774935.3806187

\vspace{3em}
\noindent\textbf{BibTeX:}
\vspace{0.5em}

\begin{Verbatim}[frame=single]
@inproceedings{10.1145/3774935.3806187,
    title       =   {Learning Behaviorally Grounded Item Embeddings via Personalized Temporal Contexts},
    author      =   {Rafael T. Sereicikas and Pedro R. Pires and Gregorio F. Azevedo and Tiago A. Almeida},
    booktitle   =   {Proceedings of the 34th ACM International Conference on User Modeling, Adaptation and
                     Personalization},
    series      =   {UMAP'26},
    location    =   {Gothenburg, Sweden},
    publisher   =   {Association for Computing Machinery},
    address     =   {New York, NY, USA},
    pages       =   {1--9},
    year        =   {2026},
    doi         =   {10.1145/3774935.3806187},
    keywords    =   {item embeddings, temporal dynamics, user behavioral modeling, representation learning,
                     time-aware recommender systems},
}
\end{Verbatim}

\newpage
\pagenumbering{arabic}

\title{Learning Behaviorally Grounded Item Embeddings via Personalized Temporal Contexts}

\author{Rafael T. Sereicikas}
    \orcid{0009-0009-9198-5469}
    \affiliation{%
        \institution{Federal University of São Carlos}
        \city{Sorocaba} 
        \state{São Paulo} 
        \country{Brazil}
    }
    \email{rafaeltofoli@estudante.ufscar.br}
    
\author{Pedro R. Pires}
    \orcid{0000-0001-7990-9097}
    \affiliation{%
        \institution{Federal University of São Carlos}
        \city{Sorocaba} 
        \state{São Paulo} 
        \country{Brazil}
    }
    \email{pedro.pires@dcomp.sor.ufscar.br}

\author{Gregorio F. Azevedo}
    \orcid{0000-0002-1096-7456}
    \affiliation{%
        \institution{Federal University of São Carlos}
        \city{Sorocaba} 
        \state{São Paulo} 
        \country{Brazil}
    }
    \email{gregorio.fornetti@estudante.ufscar.br}

\author{Tiago A. Almeida}
    \orcid{0000-0001-6943-8033}
    \affiliation{%
        \institution{Federal University of São Carlos}
        \city{Sorocaba} 
        \state{São Paulo} 
        \country{Brazil}
    }
    \email{talmeida@ufscar.br}

\renewcommand{\shortauthors}{Sereicikas et al.}

\begin{abstract}
    Effective user modeling requires distinguishing between short-term and long-term preference evolution. While item embeddings have become a key component of recommender systems, standard approaches like Item2Vec treat user histories as unordered sets (bag-of-items), implicitly assuming that interactions separated by minutes are as semantically related as those separated by months. This simplification flattens the rich temporal structure of user behavior, obscuring the distinction between coherent consumption sessions and gradual interest drifts. In this work, we introduce TAI2Vec (Time-Aware Item-to-Vector), a family of lightweight embedding models that integrates temporal proximity directly into the representation learning process. Unlike approaches that apply global time constraints, TAI2Vec is user-adaptive, tailoring its temporal definitions to individual interaction paces. We propose two complementary strategies: TAI2Vec-Disc, which utilizes personalized anomaly detection to dynamically segment interactions into semantic sessions, and TAI2Vec-Cont, which employs continuous, user-specific decay functions to weigh item relationships based on their relative temporal distance. Experimental results across eight diverse datasets demonstrate that TAI2Vec consistently produces more accurate and behaviorally grounded representations than static baselines, achieving competitive or superior performance in over 80\% of the datasets, with improvements of up to 135\%. The source code is publicly available at \url{https://github.com/UFSCar-LaSID/tai2vec}.
\end{abstract}

\ccsdesc[500]{Information systems~Recommender systems}
\ccsdesc[300]{Information systems~Personalization}
\ccsdesc[300]{Computing methodologies~Knowledge representation and reasoning}
\ccsdesc[500]{Computing methodologies~Learning latent representations}
\ccsdesc[100]{Computing methodologies~Neural networks}

\keywords{item embeddings, temporal dynamics, user behavioral modeling, representation learning, time-aware recommender systems}


\maketitle

\section{Introduction}

Understanding user behavior in recommender systems requires more than simply knowing what items a user has consumed in the past. User preferences are inherently dynamic, evolving through distinct temporal phases---ranging from rapid, goal-oriented sessions to gradual, long-term interest drifts~\cite{GrbovicECommerce2015,PiresInteract2Vec2025}. While embedding-based models have become a foundational approach for capturing semantic relationships between items~\cite{GrbovicECommerce2015,BarkanItem2Vec2016}, most standard implementations fail to account for these temporal nuances.

Originating in natural language processing with Word2Vec~\cite{MikolovDistributed2013}, the embedding paradigm was adapted to recommender systems under the assumption that items consumed by the same user are analogous to words in a sentence. Models usually learn representations by treating a user’s interaction history as a static context window. While effective at capturing global co-occurrence patterns, this approach implicitly assumes that all interactions within a user's history are equally related. Consequently, an item consumed minutes after another is treated with the same semantic weight as an item consumed months later. This ``bag-of-items'' simplification flattens the rich temporal structure of user intent.

Prior work attempted to address this limitation by incorporating temporal information into recommendation algorithms~\cite{KorenCFTemporal2009,XiangTemporal2010,VinagreOverview2015}. However, existing time-aware approaches often rely on global assumptions, e.g., applying fixed time windows or universal decay functions across the entire user base. Such heuristics fail to account for the heterogeneity of user behavior: a 24-hour gap might signal a significant session break for a high-frequency user, whereas for a casual user, it may represent mere continuity. By ignoring individual interaction paces, these models risk learning representations that are not fully grounded in the user's current preferences.

To bridge this gap, we introduce TAI2Vec (\textit{Time-Aware Item-to-Vector}), a family of lightweight embedding models that incorporates user-adaptive temporal proximity directly into the representation learning process. Unlike methods that impose arbitrary global thresholds, TAI2Vec dynamically tailors the definition of ``context'' to the individual's behavioral rhythm. We propose two complementary strategies to model this proximity:

\begin{itemize}
    \item \textbf{Discrete Segmentation (TAI2Vec-Disc):} A method that uses personal statistical thresholds to detect natural session boundaries and resample related interactions, ensuring that co-occurrence is reinforced within coherent behavioral units.
    \item \textbf{Continuous Decay (TAI2Vec-Cont):} A method that employs user-specific Z-scores to model the smooth decay of interest over time, weighting semantic relationships based on their deviation from the user's typical interaction pace.
\end{itemize}


In summary, our main contributions are:
\begin{itemize}[leftmargin=*, itemsep=2pt]
    \item[\textit{(i)}] \textbf{A User-Centric Temporal Framework:} We propose a novel perspective for embedding learning where temporal relevance is defined dynamically by the user’s individual interaction pace;
    \item[\textit{(ii)}] \textbf{Adaptive Modeling Strategies:} We introduce two time-aware mechanisms---discrete segmentation and continuous weighting---to implement temporal proximity in embedding training;
    \item[\textit{(iii)}] \textbf{Empirical Validation:} We demonstrate through extensive experiments that TAI2Vec consistently produces more accurate and behaviorally grounded representations, achieving competitive or superior performance in over 80\% of the datasets, with improvements of up to 135\% in sparse interaction scenarios.
\end{itemize}

\section{Related Work}

Traditional item embedding models learn representations based on co-occurrence statistics. Item2Vec~\cite{BarkanItem2Vec2016} is the standard in this category, treating user history as an unordered ``bag-of-items''. While effective for capturing broad semantic similarities, it ignores the sequential nature of user behavior. Similar approaches, such as Prod2Vec~\cite{GrbovicECommerce2015}, incorporate sequence ordering to capture transitions. However, these models operate on an ``ordinal'' view of time~\cite{VinagreOverview2015}, where the relative order is preserved but the magnitude of time intervals is discarded~\cite{GrbovicECommerce2015,SunBERT2019}. 

To overcome order-only models' limitations, studies have integrated explicit timestamps into recommendation logic. Extensive work has been done in RNNs and Attention-based models (e.g., TiSASRec~\cite{LiTiSASRec2020}, TALE~\cite{ParkTALE2024}), which dynamically weight interactions during the prediction phase. While powerful, they are often computationally intensive and task-specific, producing ``black-box'' predictions rather than reusable, interpretable item representations.

Within the domain of embedding learning (generating vectors that can later be reused in downstream tasks), efforts to include time have followed two main strategies:
\begin{itemize}
    \item \textit{Discrete Segmentation:} Methods like SeqI2V~\cite{PiresTime2021} and STAR \cite{YeganegiSTAR2024} introduce temporal thresholds to slice user histories into sessions (sub-contexts) and learn over these local windows.
    \item \textit{Continuous Decay:} Approaches like TimeMF~\cite{LiTimeMF2019}, TALE~\cite{ParkTALE2024}, and others~\cite{TranAttention2023,WangChorus2020} apply decay functions to down-weight older interactions, creating a smooth temporal gradient.
\end{itemize}

Despite their effectiveness, these approaches primarily incorporate temporal information at the prediction stage, rather than explicitly redefining item co-occurrence during the embedding learning process. Many of them focus on learning task-specific recommendation policies instead of producing reusable item embeddings~\cite{TranAttention2023,ParkTALE2024}, or rely on complex architectures that require substantial computational resources~\cite{CovingtonYouTube2016,LiTiSASRec2020}. More importantly, temporal modeling is typically applied globally, without adapting to individual user behaviors: discrete approaches often employ fixed temporal thresholds shared across users~\cite{PiresTime2021,YeganegiSTAR2024}, while continuous methods commonly balance interaction weights based on aggregate trends rather than user-specific temporal dynamics~\cite{WangChorus2020,ParkTALE2024}.

From a User Modeling perspective, this is suboptimal. User behavior is heterogeneous: a ``long gap'' for a high-frequency user might be a ``short gap'' for a casual user. Global parameters fail to capture these individual nuances, potentially leading to misinterpretation of user intent. TAI2Vec distinguishes itself by adopting a user-adaptive approach through individual behavioral statistics (IQR and Z-scores) to ensure the learned embeddings are grounded in the specific behavioral rhythm of each user.


\section{TAI2Vec -- Modeling Temporal Proximity}

TAI2Vec builds upon the architecture of Item2Vec~\cite{BarkanItem2Vec2016}, which 
learns representations by optimizing the probability of observing context items given a target item. However, it relies on a static definition of context: all items $j$ appearing in a user's history $H_u$ are considered equally valid context neighbors for a target item $i$, effectively setting the semantic relevance $P(j|i)$ to be uniform regardless of the duration of the time interval between interactions. This assumption discards the rich temporal signals that define user intent.

Our approach challenges this static assumption by positing that semantic co-occurrence is a function of temporal proximity. However, we argue that ``proximity'' is not an absolute metric; a one-hour gap may represent a continuity of intent for a casual user but a session break for a high-frequency user. Therefore, simply applying global time thresholds or fixed decay rates is insufficient. To capture true behavioral semantics, the definition of the context window must be user-adaptive, dynamically scaling based on the individual's historical interaction pace.

To operationalize this hypothesis, TAI2Vec weighs or filters item-item relationships during the embedding learning process, with two complementary strategies to model this adaptive proximity:
\begin{enumerate}
    \item \textbf{TAI2Vec-Disc (Discrete Segmentation):} This variant redefines the context window by identifying natural ``breaks'' in user attention. It employs distinct time thresholds, derived from the user's own inter-arrival time distribution, to resample interactions into coherent sessions.
    \item \textbf{TAI2Vec-Cont (Continuous Decay):} This variant maintains the full history but modulates the influence of context items. It applies a smooth weighting function based on the user-normalized temporal distance, prioritizing temporally dense interactions while preserving long-term context.
\end{enumerate}

\begin{figure*}[t]
    \centering
    \includegraphics[width=0.9\textwidth]{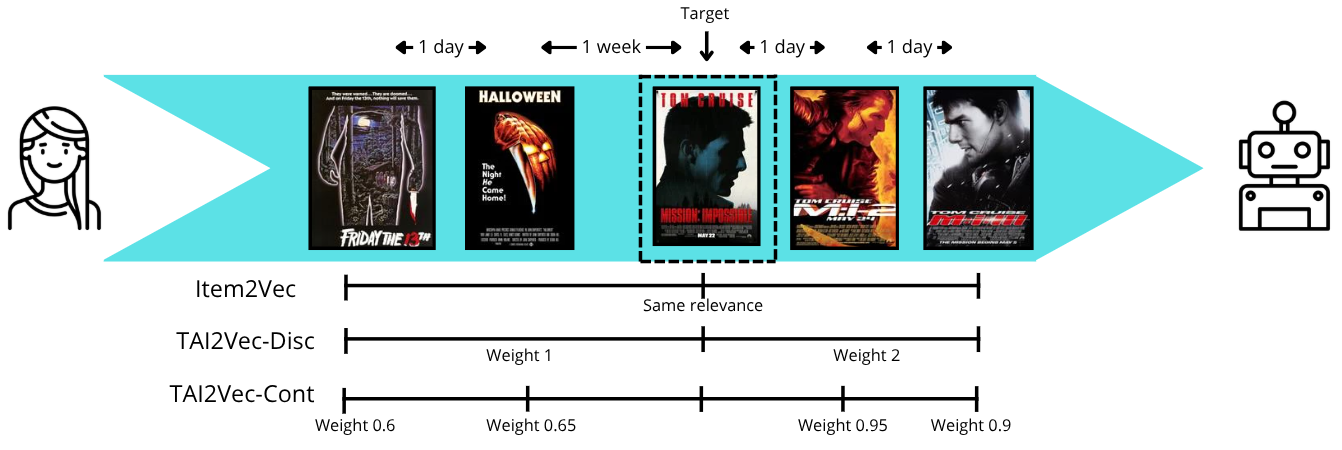}
    \caption{Visual comparison between Item2Vec and the two TAI2Vec variants. A user interaction sequence is illustrated as a sequence of consumed items annotated with their temporal intervals. Item2Vec assumes a static co-occurrence context, assigning equal importance to all items within the same user history. TAI2Vec-Disc introduces discrete temporal segmentation, creating sub-contexts that prioritize interactions within specific time windows. TAI2Vec-Cont shapes temporal proximity continuously, assigning interaction weights dynamically based on the temporal distance between items.}
    \label{fig:TAI2Vec-diagram}
    \Description{Timeline showing three recommendation models weighting a sequence of movies. Item2Vec assigns equal relevance to all films. TAI2Vec-Disc uses discrete blocks to prioritize recent interactions, splitting into different blocks when a long time span (1 week) is reached. TAI2Vec-Cont applies dynamic numerical weights, such as 0.6 to 0.95, based on the specific temporal distance between each movie and the target.}
\end{figure*}

Both variations are illustrated in Figure~\ref{fig:TAI2Vec-diagram}, in comparison with the default Item2Vec, and are explained in the following sections.

\subsection{TAI2Vec-Disc: Session Segmentation}

The TAI2Vec-Disc strategy posits that user behavior is not a continuous stream, but a sequence of distinct \textit{temporal sessions}, i.e., coherent periods of engagement separated by shifts in intent or availability. Standard methods often define sessions using fixed global thresholds (e.g., 30 minutes), which fail to account for individual interaction paces. TAI2Vec-Disc addresses this by identifying session boundaries dynamically via \textit{user-specific anomaly detection}.

We treat the detection of a session break as identifying an ``anomalous'' pause in the user's specific interaction flow. For a user $u$, we first compute the set of inter-arrival times $\Delta t_k = t_k - t_{k-1}$ for all consecutive interactions in $H_u$ (consumption history of user $u$). To prevent statistical distortions caused by near-instantaneous interactions (e.g., automated batch logging or rapid skipping), we restrict the set of intervals considered for threshold estimation to $$\mathcal{S}_u = \{ \Delta t_k \in \mathcal{T} \mid \Delta t_k > T_{\text{min}} \}$$ \noindent where $T_{\text{min}}$ is a fixed minimum valid interval.

We define the personalized session threshold $\tau_u$ using the Interquartile Range (IQR) of $\mathcal{S}_u$, a statistical measure that adapts to the user's typical cadence while avoiding outliers:

$$\tau_u = Q_3 + \lambda \cdot (Q_3 - Q_1)$$

\noindent where $Q_1$ and $Q_3$ are the first and third quartiles of the user's interval distribution, and $\lambda$ is a sensitivity hyperparameter. A new temporal session $G_u^{(m)}$ is initiated whenever $\Delta t_k > \tau_u$. This formulation ensures that a ``long gap'' is defined relatively.

Strictly separating sessions can lead to data sparsity and the loss of long-term preference signals. To mitigate this, TAI2Vec-Disc employs a \textit{dual-context} learning strategy. We retain the global co-occurrence structure but strictly up-weight interactions that occur within the same adaptive session.

We introduce a pairwise importance weight $\omega_{i,j}$ into the Skip-gram loss function. Instead of explicitly duplicating data, we define $\omega_{i,j}$ to reinforce temporally coherent pairs:

$$
    \omega_{i,j} =
    \begin{cases}
        2 & \text{if } \exists \; m \text{ s.t. } i, j \in G_u^{(m)} \quad (\text{Intra-Session}) \\
        1 & \text{otherwise} \quad \quad \quad \quad \quad \; (\text{Global Context})
    \end{cases}
$$

This effectively creates a soft constraint where short-term, temporally coherent relationships exert double the gradient influence of long-term, global co-occurrences, allowing the model to learn localized intent without forgetting the user's broader history.

\subsection{TAI2Vec-Cont: Adaptive Temporal Decay}

While TAI2Vec-Disc segments history into hard sessions, user interest drift is often gradual rather than abrupt. TAI2Vec-Cont addresses this by modeling temporal proximity as a continuous spectrum. We propose a hybrid decay mechanism that combines \textit{local behavioral statistics} with \textit{global temporal distance}.

\subsubsection{Local Decay via User-Adaptive Z-Scores}

The semantic relevance of a time interval depends on the user's typical interaction rhythm. To capture this, we employ a statistical approach that evaluates time intervals relative to the user's personal history.

To ensure consistency, we apply the $T_{min}$ noise filter in the same way as TAI2Vec-Disc, and implement an outlier treatment by clipping intervals exceeding an upper bound calculated using the IQR method, preventing long periods of inactivity from distorting the mean and standard deviation.

For each user $u$, we compute the mean $\mu_u$ and standard deviation $\sigma_u$ of their valid interval times. Additionally, to situate each interaction within the global timeline, we compute the cumulative time $t_{cum} = \sum_{j=2}^{k} \Delta t_j$. We then define the \textit{standardized temporal distance} $z_{i,j}$ between a target item $i$ and a context item $j$ as:

\begin{equation}
z_{i,j} = \frac{d_{i,j} - \mu_u}{\sigma_u + \epsilon}
\end{equation}

\noindent where $d_{i,j} = |t_{cum}^{(i)} - t_{cum}^{(j)}|$ is the absolute temporal distance in seconds. Here, $z_{i,j}$ represents how many standard deviations the interval deviates from the user's norm, effectively measuring the ``abnormality'' of the gap.

To translate this score into an importance weight, we apply a rational decay kernel controlled by a hyperparameter $\alpha$. This kernel ensures that weights decay smoothly as the interval exceeds the user's standard deviation, and guarantees that immediate interactions ($z_{i,j} \to 0$) receive near-maximum weight, while distant ones decay asymptotically to a floor $W_{\min}$:

\begin{equation}
\omega_{local}(i,j) = \max \left( W_{\min}, \ 1 - \left( \frac{z_{i,j}}{z_{i,j} + 1} \right)^\alpha \right)
\end{equation}

\begin{figure}[!htb] 
\includegraphics[width=0.4\textwidth]{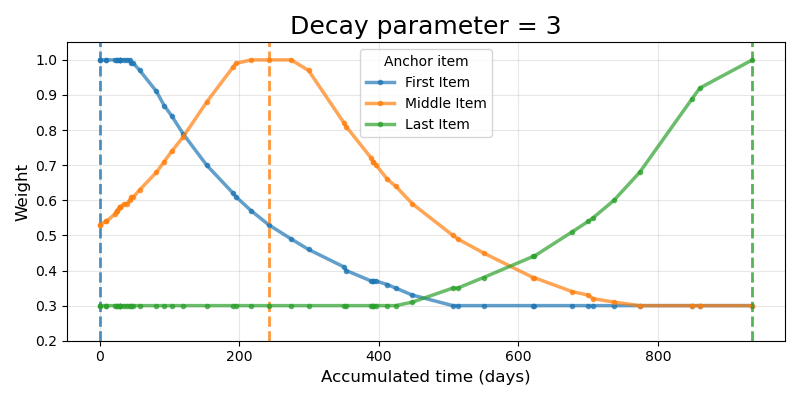} 
\includegraphics[width=0.4\textwidth]{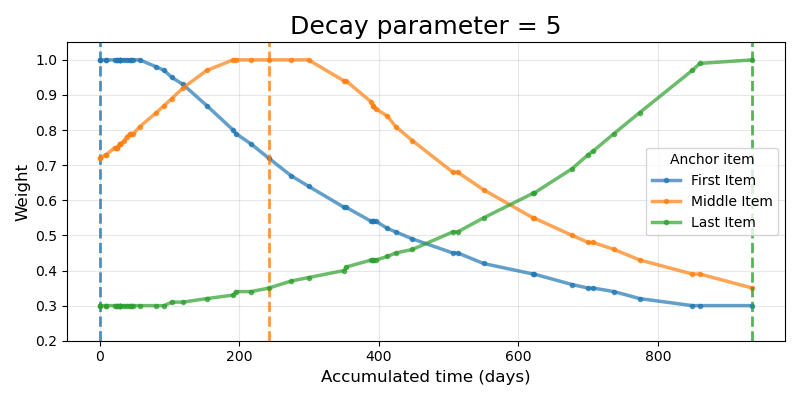} 
\caption{User-Adaptive Context (Local Decay). The weight decay curves for a single user in the Amazon Books dataset at three different points in their history. Unlike static windows, TAI2Vec-Cont dynamically adjusts the ``effective context'' based on the local density of interactions.}
\label{zscore_fig} 
\Description{Two line charts showing weight decay curves for first, middle, and last anchor items over more than 800 days. The top chart, with decay parameter 3, shows steeper curves and narrower weight distributions. The bottom chart, with decay parameter 5, shows flatter, more gradual curves where weights remain higher over longer temporal distances.}
\end{figure}

Figure~\ref{zscore_fig} illustrates the decay function for a user in the Amazon Books dataset, with $W_{min}=0.3$. The curves depict weight distributions relative to three target items: one at the start of the consumption history (blue), another in the middle (orange), and the latter at the end (green). Comparatively, increasing the decay parameter results in lighter penalization for distant interactions, whereas lower values induce a sharper transition.

\subsubsection{Global Decay via Normalized History}
Local statistics can become unstable in very short histories due to sample sparsity, or grow excessively in very long ones, leading the neural network to neglect distant preferences. To provide a stable baseline, we complement the local view with a global decay based on the item's relative position in the user's timeline.

Let $t^{(norm)}_{i} \in [0,1]$ be the min-max normalized timestamp of interaction $i$. The global weight is defined as a linear decay function, which penalizes interactions that are distant in the absolute timeline, regardless of the local interaction density:

\begin{equation}
\omega_{global}(i,j) = 1 - (1 - W_{\min}) \cdot | t^{(norm)}_{i} - t^{(norm)}_{j} |
\end{equation}

\begin{figure}[!htb]
\includegraphics[width=0.4\textwidth]{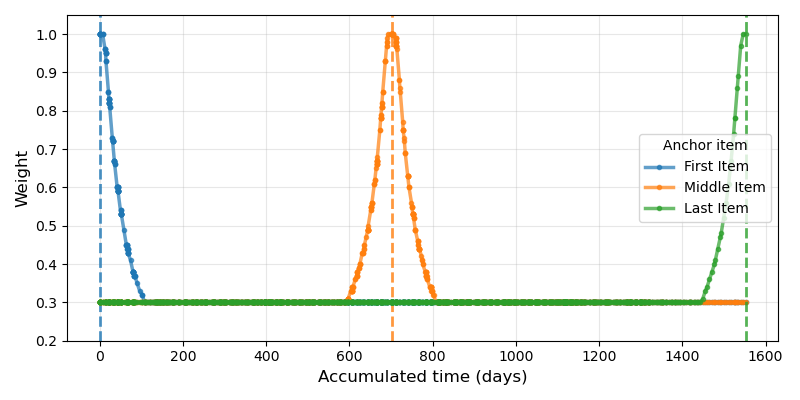}
\includegraphics[width=0.4\textwidth]{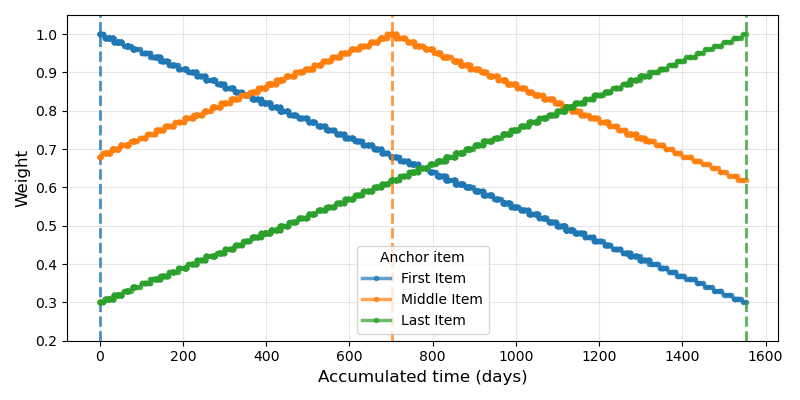}
\caption{Mitigating History Bias (Local vs. Global). A comparison of decay strategies for a user with a 5-year history. The Local approach (top panel) risks dropping long-term signals too aggressively. The Global approach (bottom panel) maintains a stable, linear baseline.}
\label{linear_curve}
\Description{Two line charts comparing Local and Global decay strategies over a 1600-day history. The top panel shows Local decay with sharp, narrow bell curves centered around anchor items, where weights drop rapidly toward 0.3. The bottom panel shows Global decay with linear shapes, maintaining significantly higher weights and more stable signals across the entire timeline. Both charts track first, middle, and last anchor items to illustrate how the model mitigates history bias.}
\end{figure}

Figure \ref{linear_curve} illustrates how linear modeling offers a complementary perspective. The example depicts three items consumed by a user with an unusually extensive history for the Amazon Books dataset (1,700 records over 5 years). Here, the limitation of the local approach (top panel) is evident: the minimum weight is reached for interactions merely ~100 days distant from the target. The global approach (bottom panel), however, considers the relative distance across the entire trajectory, preventing the premature discarding of the user's long-term preferences.

\subsubsection{Unified Weighting Strategy}
The final weight $\omega_{i,j}$ used to scale the gradient in the embedding loss function is the arithmetic mean of the local and global perspectives:

\begin{equation}
\omega_{i,j} = \frac{\omega_{local}(i,j) + \omega_{global}(i,j)}{2}
\label{eq:unified_weight}
\end{equation}


\begin{figure}[!htb]
\includegraphics[width=0.4\textwidth]{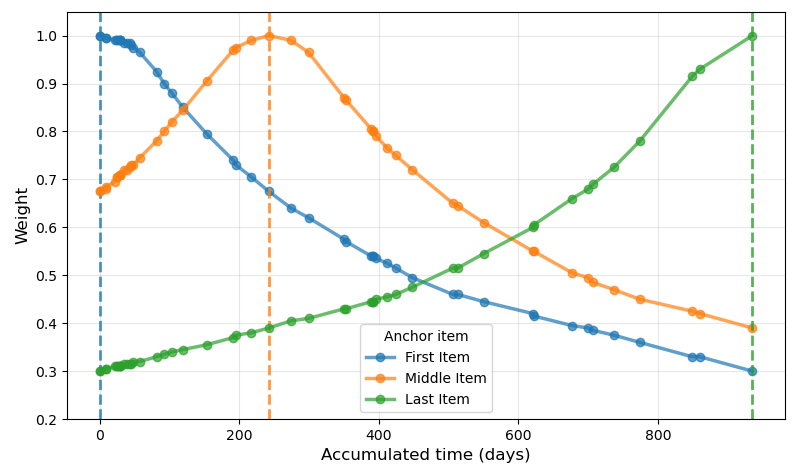}
\caption{Unified Temporal Weighting. The final hybrid weight $\omega_{i,j}$ balances the user's specific interaction pace with global stability. This creates a flexible context window that filters noise from distant interactions while retaining sufficient signal for representation learning.}
\label{final_curve}
\Description{Line chart representing a Unified Temporal Weighting strategy, which merges local and global approaches. Three smooth, bell-shaped curves represent anchor items over 1000 days. The blue curve decays from the start, the orange curve peaks centrally, and the green curve ascends toward the end.}
\end{figure}

The resulting hybrid weight, depicted in Figure~\ref{final_curve}, balances the sensitivity of the local user clock with the stability of the global timeline. By integrating $\omega_{i,j}$ as a multiplier in the loss function, TAI2Vec-Cont dynamically focuses the learning process on temporally relevant pairs, attenuating the noise from distant interactions without sacrificing the semantic richness of the user's full history.

\subsection{User Representation and Recommendation}

Upon convergence, the model yields a semantic embedding matrix $\mathbf{V} \in \mathbb{R}^{|\mathcal{I}| \times d}$, where each row $\mathbf{v}_i$ represents an item $i$ in the latent space. A key advantage of TAI2Vec is that while the training process is temporally complex, the resulting embeddings intrinsically encode these temporal structures. This allows us to use efficient, standard aggregation techniques for inference without requiring the heavy sequential computation typical of RNNs or Transformers.

We derive a user's latent representation $\mathbf{u}$ dynamically by aggregating their interaction history $H_u$, as the centroid of the items they have consumed, following standard Item2Vec protocols:

\begin{equation}
\mathbf{u} = \frac{1}{|H_u|} \sum_{j \in H_u} \mathbf{v}_j
\label{eq:user_vector}
\end{equation}

For a candidate item $i$ (where $i \notin H_u$), the preference score $S_{u,i}$ is computed as the cosine similarity between the user vector and the candidate item embedding. This is equivalent to the average similarity between the candidate and all historical items:

\begin{equation}
S_{u,i} = \mathbf{u}^\top \mathbf{v}_i = \frac{1}{|H_u|} \sum_{j \in H_u} \mathbf{v}_j^\top \mathbf{v}_i
\label{eq:score}
\end{equation}

\begin{figure}[!htb]
\includegraphics[width=0.4\textwidth]{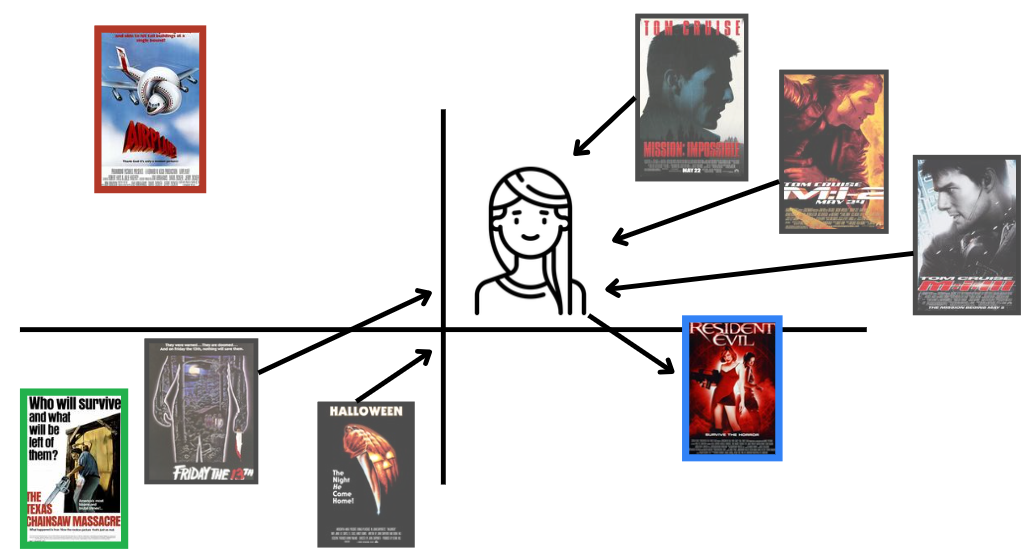}
\caption{Illustration of user modeling via item embeddings. A user is implicitly represented as the aggregation of previously consumed items in the latent space. Candidate items, with a highlighted border, are scored by their average proximity to the consumed items, and the most aligned with the user's interaction profile is selected for recommendation.}
\label{fig:item_item_rec}
\Description{A plot illustrating user modeling via item embeddings in latent space. Movie posters represent consumed and candidate items. A user is represented by an avatar in the center of the plot. Consumed items have arrows pointing to the user. Candidate items have highlighted borders. The recommended item has an arrow pointing from the user to it, and it is the item with the most proximity to the consumed items in the plot.}
\end{figure}

Finally, we generate recommendations by creating a ranking of all non-consumed items in descending order of $S_{u,i}$ and selecting the top-$K$ candidates. This approach ensures that recommendation quality is driven by temporally grounded semantics learned during training. Figure~\ref{fig:item_item_rec} illustrates the intuition behind this implicit user representation and the resulting item-to-item recommendation.

\section{Experimental setup}

We conducted multiple recommendation tasks to evaluate the effectiveness of TAI2Vec variations and assess how incorporating user-adaptive temporal modeling during the representation learning phase impacts downstream recommendation quality.

\subsection{Datasets and Preprocessing}

The experiments were executed over a diverse set of eight publicly available datasets widely used in recommender systems research. These datasets were selected to cover a diverse range of domains (i.e., different e-commerce categories and movies) and interaction densities, allowing us to test the model's adaptability to different user behavioral rhythms. Table~\ref{tab:dados} summarizes their statistics.


\begin{table}[!htpb]
\centering
\caption{Dataset Statistics. Columns $|U|$, $|I|$, and $|R|$ contain, respectively, the number of users, items, and interactions. Additionally, the Period column indicates the total timespan of the dataset in months, and $\overline{\Delta t}$ represents the average time difference between consecutive user interactions in days.}
\label{tab:dados}
\resizebox{0.48\textwidth}{!}{%
\begin{tabular}{lcccccc}
\hline
\textbf{Dataset} & $\boldsymbol{|U|}$ & $\boldsymbol{|I|}$ & $\boldsymbol{|R|}$ & \textbf{Period} & $\boldsymbol{\overline{\Delta t}}$ \\
\hline
\textbf{Amazon-Beauty}\footnotemark[1]   &   631,986 &   112,565 &   693,929 & 234.00 & 422.9 \\
\textbf{Amazon-Books}\footnotemark[1]    & 1,008,954 &   206,710 & 2,115,793 & 195.36 & 485.7 \\
\textbf{Amazon-Games}\footnotemark[1]    &   757,210 &    80,316 & 2,579,662 & 204.08 & 241,83 \\
\textbf{BestBuy}\footnotemark[2]         & 1,268,702 &    69,858 & 1,862,782 & 2.64 & 3.7 \\
\textbf{CiaoDVD}\footnotemark[3]         &    16,923 &    14,800 &    65,038 & 151.64 & 81.4 \\
\textbf{MovieLens-100k}\footnotemark[4]  &       943 &     1,682 &   100,000 & 7.08 & 0.4 \\
\textbf{MovieLens-1M}\footnotemark[4]    &     6,039 &     3,628 &   836,478 & 34.08 & 1.1 \\
\textbf{RetailRocket}\footnotemark[5]    &     1,344&     2,158 &   6,513 & 4,56 & 4.64 \\
\hline
\end{tabular}%
}
\end{table}

\footnotetext[1]{Amazon Reviews'23. Available at: \url{https://amazon-reviews-2023.github.io} (visited on \today)}
\footnotetext[2]{Data Mining Hackathon on BIG DATA (7GB). Available at: \url{www.kaggle.com/c/acm-sf-chapter-hackathon-big} (visited on \today)}
\footnotetext[3]{CiaoDVD. Available at: \url{https://guoguibing.github.io/librec/datasets.html} (visited on \today)}
\footnotetext[4]{MovieLens | GroupLens. Available at: \url{https://grouplens.org/datasets/movielens/} (visited on \today)}
\footnotetext[5]{Retailrocket recommender system dataset. Available at: \url{kaggle.com/retailrocket/ecommerce-dataset} (visited em \today.)}

To ensure a proper context for embedding learning, we filtered users and items with fewer than 5 interactions, removed duplicate user-item pairs, and binarized the explicit ratings, considering only interactions that exceeded a midpoint value in the rating range.




\subsection{Experimental Protocol}

\paragraph{Temporal Splitting.} To strictly prevent temporal data leakage, we employed a global temporal split strategy~\cite{GusakTime2025}. For each dataset, all interactions were ordered chronologically. We partitioned the data into Training (first 80\%), Validation (next 10\%), and Test (last 10\%) sets based on absolute timestamps. This ensures that the model simulates a real-world deployment scenario.

\paragraph{Handling Cold-Start.} Users or items appearing in the Validation or Test sets but not in the Training set were removed, as the benchmarked embedding-based methods cannot generate representations for unseen entities without auxiliary content.

\subsection{Baselines and Hyperparameters}

We benchmarked our proposed variants against standard Item2Vec~\cite{BarkanItem2Vec2016}, a widely adopted lightweight model for static, co-occurrence-based item embedding, and SeqI2Vec~\cite{PiresTime2021}, a skip-gram model with global and non-adaptive temporal dynamics. 

\paragraph{Hyperparameter Tuning.} We performed a grid search over the Validation set to identify optimal configurations:
\begin{itemize}
    \item Context Window ($w$): $w=\{5, 10\}$ on baseline Item2Vec to check if static context length alone drives performance; and $w=10$ on time-based models to isolate temporal effects.
    \item Embedding Dimension ($d$): 50.
    \item Negative Sampling ($k$): 7 negative samples using a sampling exponent $\eta \in \{-1, -0.5, 0.5, 1\}$.
    \item Subsample: Subsampling threshold $t \in \{10^{-3}, 10^{-4}, 10^{-5}\}$
    \item Optimization: Adam optimizer with a batch size of $10^{14}$ and learning rate $ \alpha \in \{0.25, 0.025\}$.
    \item Epochs: Fixed epochs with linear decay $e \in \{20, 50, 100\}$.
\end{itemize}

For the specific temporal parameters of TAI2Vec:

\begin{itemize}
    \item TAI2Vec-Disc: Sensitivity $\lambda \in \{1.0, 1.5, 2.0\}$.
    \item TAI2Vec-Cont: Decay rate $\alpha \in \{3.0, 5.0\}$ and minimum weight $W_{min} \in \{0.3, 0.5\}$.
    \item Noise Filter: A minimum interval filter $T_{min} = 300$ seconds was applied to ignore system-generated rapid logs.
\end{itemize}

After tuning, the Training and Validation sets were merged to train the final models, which were then evaluated on the Test set.

\section{Results and Discussions}

To evaluate the proposed TAI2Vec variants, we conducted experiments according to the aforementioned evaluation protocol, addressing the following research questions (RQs):

\begin{itemize}
    \item \textbf{RQ1 (Efficacy):} Does modeling temporal proximity yield more behaviorally grounded item representations compared to static co-occurrence models?
    \item \textbf{RQ2 (Personalization):} How does a user-adaptive temporal strategy compare to a more general, global approach?
    \item \textbf{RQ3 (Strategy):} How do discrete segmentation (TAI2Vec-Disc) and continuous decay (TAI2Vec-Cont) compare in capturing different user behavioral patterns?
\end{itemize}

To answer these RQs, we report the Normalized Discounted Cumulative Gain (NDCG) and the Root Mean Square Error (RMSE), computed over the test set. 
NDCG accounts for the item's position, rewarding algorithms that place true positives higher in the ranking, while RMSE estimates how well the representation performs on score prediction tasks. For the former, higher values indicate better recommendation performance, while for the latter, lower values are preferred. To support transparent and reproducible research, we have released the full source code used in our experiments, at \url{https://github.com/UFSCar-LaSID/tai2vec}. The repository also includes additional analysis, such as results for the Hit Rate and alternative data splits.





\begin{figure*}[!htb]
    \centering
    \includegraphics[width=1\textwidth]{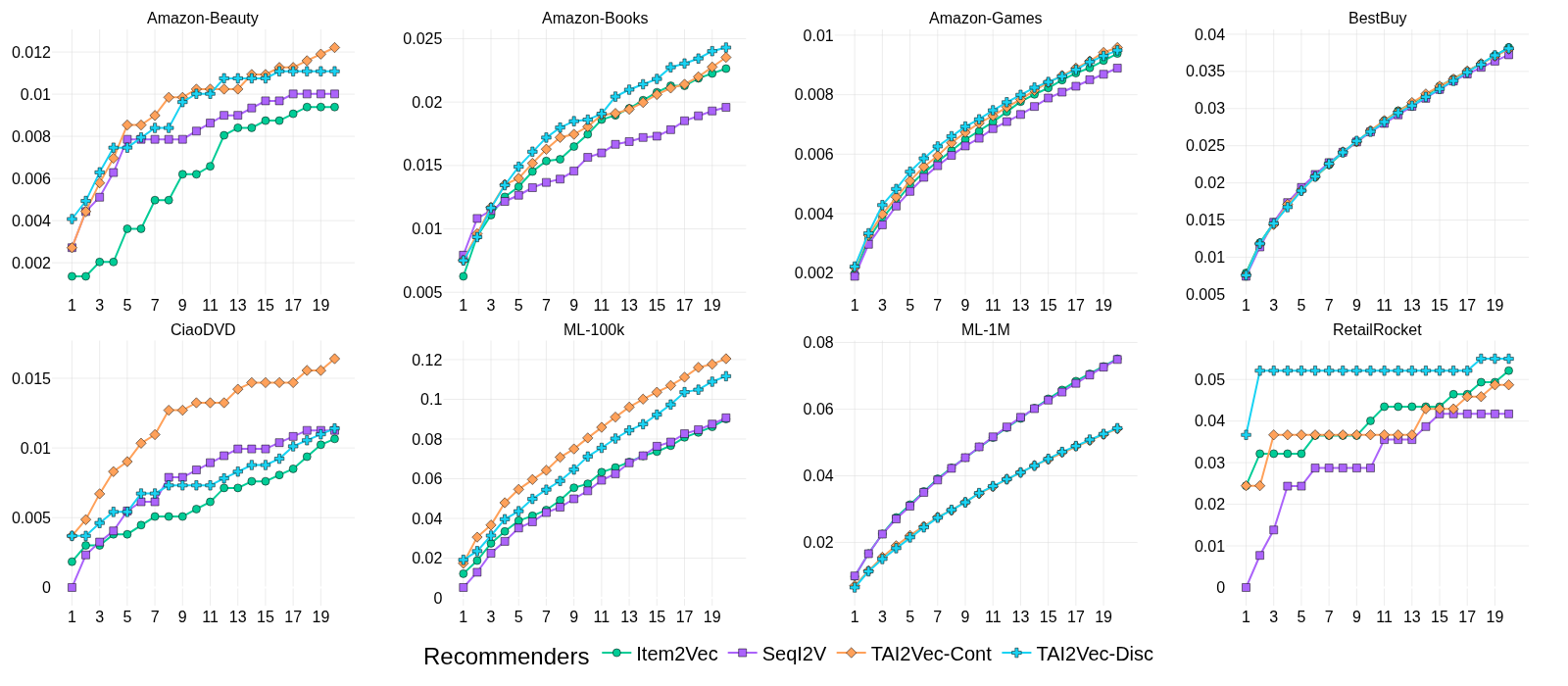}
    \caption{Robustness across List Sizes. Average NDCG@$N$ results for multiple values of $N$ across different datasets. TAI2Vec variants (orange/purple) consistently outperform the static baseline (green) across ranking depths, confirming that the improvement is not an artifact of a specific cutoff.}
    \label{NDCG_results}
    \Description{A grid of eight line charts comparing four recommendation models—Item2Vec, SeqI2V, TAI2Vec-Cont, and TAI2Vec-Disc—across various datasets, including Amazon-beauty, Amazon-books, Amazon-games, BestBuy, CiaoDVD, MovieLens-100k, MovieLens-1M, and RetailRocket. Each chart plots Average NDCG@N against list sizes ranging from 1 to 20. The TAI2Vec variants consistently show higher performance curves than the Item2Vec baseline across different ranking depths.}
\end{figure*}

Table~\ref{tab:ndcg_merged} presents the results for the top-$10$ recommendations. To facilitate comparison, darker cells indicate superior results. The proposed TAI2Vec models demonstrate consistent superiority over the static Item2Vec baseline in 7 out of 8 datasets, providing clear evidence for \textbf{RQ1}. The magnitude of improvement is strongly correlated with data sparsity. In Amazon-Beauty and CiaoDVD---datasets characterized by sparse, irregular interaction intervals---TAI2Vec-Cont achieves relative gains of 65.16\% and 135.52\%. This confirms that when user signals are fragmented, explicitly modeling temporal proximity is crucial for recovering meaningful relationships. Additionally, to ensure our findings are robust to the choice of list size, we visualize performance trends for $N \in [1,20]$ in Figure~\ref{NDCG_results}.

\begin{table}[ht]
\centering
\caption{NDCG@10 with percentage improvement over Item2Vec (in parentheses).}
\label{tab:ndcg_merged}
\resizebox{0.48\textwidth}{!}{
\begin{tabular}{lcccc}
\toprule
Dataset & Item2Vec & SeqI2V & TAI2Vec-Cont & TAI2Vec-Disc \\
\midrule
Amz-Beauty & \cellcolor[gray]{0.95}0.0062 
& \cellcolor[gray]{0.7216}0.0083 (+33.07\%) 
& \cellcolor[gray]{0.5}0.0102 (+65.16\%) 
& \cellcolor[gray]{0.5247}0.0100 (+61.58\%) \\

Amz-Books & \cellcolor[gray]{0.6751}0.0175 
& \cellcolor[gray]{0.95}0.0156 (-10.40\%) 
& \cellcolor[gray]{0.5834}0.0181 (+3.47\%) 
& \cellcolor[gray]{0.5}0.0186 (+6.62\%) \\

Amz-Games & \cellcolor[gray]{0.7833}0.0068 
& \cellcolor[gray]{0.95}0.0065 (-3.49\%) 
& \cellcolor[gray]{0.593}0.0070 (+3.99\%) 
& \cellcolor[gray]{0.5}0.0072 (+5.93\%) \\

BestBuy & \cellcolor[gray]{0.6857}0.0270 
& \cellcolor[gray]{0.95}0.0269 (-0.42\%) 
& \cellcolor[gray]{0.5}0.0271 (+0.30\%) 
& \cellcolor[gray]{0.8876}0.0269 (-0.32\%) \\

CiaoDVD & \cellcolor[gray]{0.95}0.0056 
& \cellcolor[gray]{0.7842}0.0084 (+49.92\%) 
& \cellcolor[gray]{0.5}0.0132 (+135.52\%) 
& \cellcolor[gray]{0.8497}0.0073 (+30.21\%) \\

ML-100k & \cellcolor[gray]{0.8898}0.0575 
& \cellcolor[gray]{0.95}0.0539 (-6.20\%) 
& \cellcolor[gray]{0.5}0.0805 (+40.17\%) 
& \cellcolor[gray]{0.6583}0.0712 (+23.86\%) \\

ML-1M & \cellcolor[gray]{0.5001}0.0487 
& \cellcolor[gray]{0.5}0.0487 (+0.01\%) 
& \cellcolor[gray]{0.95}0.0347 (-28.81\%) 
& \cellcolor[gray]{0.9445}0.0348 (-28.46\%) \\

RetailRocket & \cellcolor[gray]{0.7318}0.0401 
& \cellcolor[gray]{0.95}0.0287 (-28.32\%) 
& \cellcolor[gray]{0.7968}0.0367 (-8.43\%) 
& \cellcolor[gray]{0.5}0.0521 (+30.09\%) \\

\bottomrule
\end{tabular}
}
\end{table}

When comparing SeqI2V with TAI2Vec---particularly its discrete variant, which shares a temporal segmentation strategy---our results highlight the advantage of adopting a user-adaptive perspective, directly addressing \textbf{RQ2}. While SeqI2V incorporates temporal information uniformly across users, TAI2Vec dynamically adjusts it to individual behavioral patterns. Empirically, TAI2Vec variants outperform SeqI2V in 88\% of the datasets, and when considering improvements over the static Item2Vec baseline, SeqI2V achieves an average gain of 4.27\%, whereas TAI2Vec-Disc reaches 16.19\%, representing up to a fourfold increase in relative improvement.

For the NDCG, MovieLens-1M is the only dataset where Item2Vec and SeqI2V outperform our variants. This exception highlights the boundary conditions of our approach. ML-1M differs fundamentally from the others: it has high interaction density within a relatively short span, and importantly, its timestamps often reflect rating time (retrospective) rather than consumption time (real-time). In such dense, stable environments, the ``global context'' is already robust, and aggressive temporal decay may filter out valid co-occurrences. This suggests TAI2Vec is best suited for real-world scenarios with temporal heterogeneity, rather than curated, dense rating datasets.

Regarding \textbf{RQ3}, we can compare the two strategies to reveal trade-offs based on domain characteristics: TAI2Vec-Cont emerges as the most robust variant overall; its smooth decay function is better at capturing gradual preference drifts, which are common in e-commerce. TAI2Vec-Disc performs best on Amazon-Books and MovieLens-100k, domains which often involve distinct ``sessions'' (e.g., reading a book series or a weekend movie binge), and where the IQR-based segmentation successfully isolates these coherent units, preventing noise from unrelated sessions.


In addition to NDCG, Table~\ref{tab:rmse} reports the performance in terms of RMSE, providing a complementary perspective on the accuracy of the embeddings. The results reveal a more nuanced behavior compared to ranking metrics. TAI2Vec variations achieve the best performance in 62\% of the datasets, indicating that incorporating temporal proximity can also improve score prediction tasks. These findings are consistent with \textbf{RQ1}, reinforcing that time-aware and user-centric approaches contribute not only to ranking quality but also to more accurate preference estimation.

\begin{table}
\centering
\caption{Comparison for RMSE with Grayscale}
\label{tab:rmse}
\resizebox{0.48\textwidth}{!}{\begin{tabular}{lcccc}
\toprule
Dataset & Item2Vec & SeqI2V & TAI2Vec-Cont & TAI2Vec-Disc \\
\midrule
Amz-Beauty & \cellcolor[gray]{0.95}1.8255 & \cellcolor[gray]{0.8242}1.8123 & \cellcolor[gray]{0.5}1.7782 & \cellcolor[gray]{0.522}1.7805 \\
Amz-Books & \cellcolor[gray]{0.5801}1.6123 & \cellcolor[gray]{0.5}1.6110 & \cellcolor[gray]{0.8533}1.6169 & \cellcolor[gray]{0.95}1.6185 \\
Amz-Games & \cellcolor[gray]{0.6767}1.6773 & \cellcolor[gray]{0.8909}1.6963 & \cellcolor[gray]{0.5}1.6616 & \cellcolor[gray]{0.95}1.7016 \\
BestBuy & \cellcolor[gray]{0.9472}0.4996 & \cellcolor[gray]{0.652}0.4991 & \cellcolor[gray]{0.5}0.4988 & \cellcolor[gray]{0.95}0.4996 \\
CiaoDVD & \cellcolor[gray]{0.6627}0.4992 & \cellcolor[gray]{0.7506}0.4997 & \cellcolor[gray]{0.5}0.4982 & \cellcolor[gray]{0.95}0.5010 \\
ML-100k & \cellcolor[gray]{0.5554}1.2323 & \cellcolor[gray]{0.5}1.2242 & \cellcolor[gray]{0.7992}1.2679 & \cellcolor[gray]{0.95}1.2899 \\
ML-1M & \cellcolor[gray]{0.5535}1.0568 & \cellcolor[gray]{0.5}1.0510 & \cellcolor[gray]{0.9022}1.0945 & \cellcolor[gray]{0.95}1.0997 \\
RetailRocket & \cellcolor[gray]{0.6816}0.4672 & \cellcolor[gray]{0.95}0.5138 & \cellcolor[gray]{0.5}0.4356 & \cellcolor[gray]{0.584}0.4502 \\
\bottomrule
\end{tabular} }
\end{table}

Mirroring the trend observed in NDCG, in denser datasets such as ML-1M, TAI2Vec variants underperform compared to Item2Vec and SeqI2V. This behavior highlights that, in scenarios with stable and abundant interactions, static co-occurrence signals may already provide sufficient information for accurate score estimation.

Overall, these results suggest that TAI2Vec maintains competitive performance across both ranking and prediction tasks, while particularly excelling in realistic scenarios where temporal dynamics play a central role in shaping user behavior.

\subsection{Statistical Analysis}

To better understand the observed performance differences, we conducted non-parametric statistical tests comparing the TAI2Vec variants with the baseline methods.

First, we applied the Friedman test, evaluated using the chi-square approximation. The test did not reveal statistically significant differences among the compared methods ($\chi^2(3) = 5.96$, $p = 0.113$). However, the associated effect size, measured using Kendall's $W = 0.24$, indicates a small-to-moderate level of agreement in performance differences across datasets. This suggests the presence of a consistent, albeit not statistically strong, trend favoring the proposed time-aware variants.

We further conducted pairwise comparisons using the Wilcoxon signed-rank test. The results are reported in Table~\ref{tab:wilcoxon-test}, including $p$-values and effect sizes ($r$), along with their standard magnitude interpretation (\textbf{S}mall, \textbf{M}edium, \textbf{L}arge).

\begin{table}[hptb]
\centering
\caption{Results of the Wilcoxon signed-rank test}
\label{tab:wilcoxon-test}
\begin{tabular}{cll}
    \toprule
    & TAI2Vec-Cont & TAI2Vec-Disc \\
    \midrule
     Item2Vec   & $[p = 0.38 \text{ | } r = 0.31 \text{ (\textbf{\texttt{M}})}]$ & $[p = 0.25 \text{ | } r = 0.41 \text{ (\textbf{\texttt{M}})}]$ \\
     SeqI2V     & $[p = 0.15 \text{ | } r = 0.51 \text{ (\textbf{\texttt{L}})}]$ & $[p = 0.30 \text{ | } r = 0.39 \text{ (\textbf{\texttt{M}})}]$ \\
     \bottomrule
\end{tabular}
\end{table}

The Wilcoxon tests did not reject the null hypothesis in any pairwise comparison, indicating that the observed differences are not statistically significant at the $\alpha = 0.05$ level. Nevertheless, all comparisons exhibit at least medium effect sizes, with a large effect observed when comparing TAI2Vec-Cont and SeqI2V ($r = 0.51$).

Taken together, these findings suggest a consistent empirical trend in favor of the proposed methods, which is also reflected in the descriptive results, where TAI2Vec variants outperform the baseline in 87\% of the evaluated settings with an average relative gain exceeding 20\% when combined. The results highlight the importance of incorporating temporal proximity in a user-adaptive strategy for learning behaviorally grounded embeddings.

\subsection{Ablation Study}

\begin{figure}[!hptb]
    \centering
    \includegraphics[trim={2cm 0.5cm 2cm 3.5cm},clip,width=0.5\textwidth]{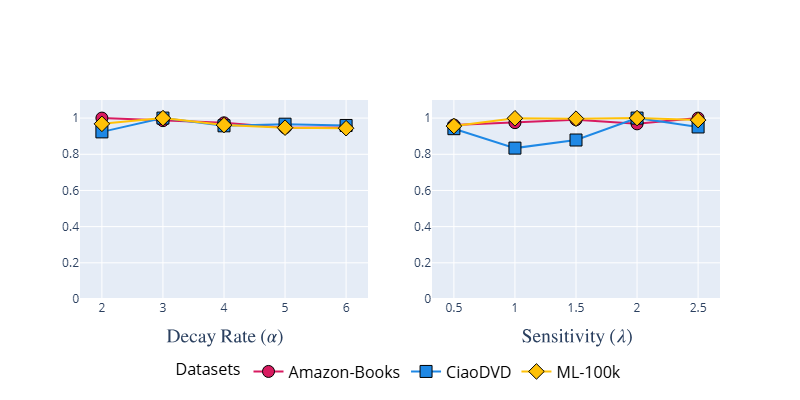}
    \caption{Scaled NDCG across varying temporal hyperparameters of TAI2Vec.}
    \label{fig:ablation_study}
    \Description{Two side-by-side line charts showing the Scaled NDCG performance of TAI2Vec across different temporal hyperparameters for three datasets: Amazon-Books, CiaoDVD, and ML-100k. The left chart shows stability as the Decay Rate (alpha) varies from 2 to 6. The right chart shows performance fluctuations as Sensitivity (lambda) varies from 0.5 to 2.5, with the ML-100k dataset maintaining the highest relative stability across both parameters.}
\end{figure}

We evaluated the sensitivity of the proposed models to their temporal hyperparameters: the decay rate $\alpha$ for TAI2Vec-Cont and the sensitivity $\lambda$ for TAI2Vec-Disc. For each parameter, we varied its value while fixing all other hyperparameters, and measured performance on three representative datasets (CiaoDVD, ML-100k, and Amazon-Books). To enable cross-dataset comparison, results were normalized using Min–Max scaling, and are shown in Figure~\ref{fig:ablation_study}.

Overall, both parameters exhibit limited influence on performance, with relatively stable behavior across the explored ranges. This is reflected in the low standard deviation of the normalized scores (0.0244 for $\alpha$ and 0.0331 for $\lambda$), indicating that variations in these parameters lead to only minor fluctuations in recommendation quality. In particular, performance curves remain largely flat across the tested values, suggesting the absence of a narrow optimal region and reinforcing the robustness of the proposed models. 

Among the two, $\alpha$ tends to yield better results at lower values, a trend consistently observed in the Amazon-Books and ML-100k datasets. In contrast, no consistent pattern is observed for $\lambda$, which displays slightly higher variability and stronger dataset dependency. This behavior is expected, as TAI2Vec-Disc relies on discrete session segmentation, where small changes in the threshold can induce abrupt structural differences in the training data.

These findings suggest that the proposed temporal mechanisms are robust to hyperparameter selection, requiring only coarse tuning to achieve competitive performance.

\section{Conclusions and Future Work}

In this work, we challenged the prevailing ``bag-of-items'' assumption in lightweight representation learning, arguing that semantic relevance is intrinsically tied to the temporal rhythm of user interaction. We introduced TAI2Vec, a user-adaptive embedding framework that treats the context window as a dynamic behavioral unit. By operationalizing temporal proximity through discrete session segmentation (TAI2Vec-Disc) and continuous statistical decay (TAI2Vec-Cont), we demonstrated that item embeddings can be effectively grounded in the reality of user consumption patterns.

Our empirical evaluation across eight diverse datasets confirms that incorporating user-specific temporal dynamics significantly enhances recommendation quality. TAI2Vec consistently outperformed the baselines in over 80\% of the evaluated settings, with gains exceeding 100\% in sparse, irregular environments like CiaoDVD. These findings suggest that for users with fragmented histories, the relative time between actions is a far more critical signal of intent than mere co-occurrence.

Looking forward, we envision four key directions to expand this user-centric modeling:
\begin{enumerate}
    \item \textit{Online and Incremental Learning:} Integrating TAI2Vec into streaming recommendation frameworks~\cite{SilvaMulti2022}. Since our decay functions are computed dynamically, they naturally support non-stationary environments where user preferences drift in real-time;
    \item \textit{Explainability and Visualization:} Unlike black-box sequential models, TAI2Vec's decay curves (e.g., Z-scores) offer an interpretable metric of ``current user interest''. We plan to explore how these temporal weights can be surfaced to users or analysts to explain \textit{why} a recommendation was made (e.g., ``Because this item fits your current 2-hour session'');
    \item \textit{Hybrid Architectures:} Exploring unified frameworks that combine the sharp boundary detection of our discrete model with the smooth drift modeling of our continuous variant, potentially using attention mechanisms to learn the optimal balance for each user automatically;
    \item \textit{Content-Enriched Embeddings:} Incorporating item-side metadata to augment TAI2Vec representations, enabling hybrid modeling and improving robustness in cold-start scenarios where interaction data is scarce.
\end{enumerate}

\section{Ethical Considerations and Broader Impact}
\label{sec:ethics}

As recommender systems rely on behavioral data, it is imperative to address the ethical implications of modeling user-specific dynamics.

\paragraph{Privacy and Behavioral Fingerprinting.}
TAI2Vec leverages temporal interaction patterns, which may inadvertently expose sensitive behavioral traits. To mitigate this risk, the model operates on relative time intervals and encodes them as abstract semantic weights, reducing the possibility of reconstructing raw user activity. In practical deployments, additional safeguards such as noise injection into user-specific statistics can further enhance privacy.

\paragraph{Temporal Bias and Filter Bubbles.}
By prioritizing temporally coherent interactions, TAI2Vec may overemphasize short-term user intent, potentially limiting exposure to long-term preferences. This can lead to a form of temporal bias, where recommendations become overly focused on recent behavior. To address this, temporal modeling should be complemented with diversity-aware strategies.

\paragraph{Environmental Impact and Accessibility.}
TAI2Vec is designed as a lightweight extension of Item2Vec, maintaining a low computational footprint while incorporating temporal information. This efficiency reduces energy consumption compared to deep sequential models and makes time-aware recommendations more accessible to systems with limited computational resources.

\begin{acks}
This study was financed, in part, by the Brazilian Agencies CNPq (grant \#311867/2023-5), CAPES (Finance Code 88887.854357/2023-00), and FAPESP (Process Numbers \#2021/14591-7, \#2023/00158-5, and \#2024/15919-4). Limited use of generative AI tools was made to assist with language refinement. The authors are solely responsible for all content of this paper.
\end{acks}

\balance
\bibliographystyle{ACM-Reference-Format}
\bibliography{references}


\end{document}